\documentclass{article}
\usepackage[utf8]{inputenc} 
\usepackage[T1]{fontenc}
\usepackage{spconf,amsmath,graphicx}
\usepackage{enumitem}
\setlist{nosep, leftmargin=14pt}
\usepackage{mwe} 
\usepackage{hyperref}       
\usepackage{url}            
\usepackage{booktabs}       
\usepackage{amsfonts}       
\usepackage{nicefrac}       
\usepackage{microtype}      
\usepackage{lipsum}
\usepackage{stackengine}
\usepackage{float}
\usepackage{amssymb}
\usepackage{subfig}
\usepackage{float}
\usepackage{color, colortbl}
\usepackage{rotating}
\usepackage{subfig}
\usepackage{tikz}
\usepackage{xcolor}
\usepackage{animate}
\usepackage{mathrsfs}

\usepackage[ruled,vlined]{algorithm2e}    
\newcommand{\vect}[1]{\boldsymbol{#1}}

\hypersetup{
    colorlinks,
    linkcolor={red!50!black},
    citecolor={blue!80!black},
    urlcolor={blue!80!black}
}

\title{The Dynamics of Triple Interactions in Resting fMRI: \\ Insights into Psychotic Disorders}
\name{Qiang Li$^{1}$\thanks{qli27@gsu.edu \\ \textbf{\textit{IEEE International Symposium on Biomedical Imaging (ISBI 2025)}}}, Vince D. Calhoun$^{1,2}$, Armin Iraji$^{1,2}$}
\address{$^{1}$Tri-Institutional Center for Translational Research in Neuroimaging and Data Science (TReNDS), \\
Georgia State, Georgia Tech, and Emory University, Atlanta, GA 30303, United States \\
$^{2}$Department of Computer Science, Georgia State University, Atlanta, GA 30303, United States}

\begin{document}
\maketitle

\begin{abstract}
The human brain dynamically integrated and configured information to adapt to the environment. To capture these changes over time, dynamic second-order functional connectivity was typically used to capture transient brain patterns. However, dynamic second-order functional connectivity typically ignored interactions beyond pairwise relationships. To address this limitation, we utilized dynamic triple interactions to investigate multiscale network interactions in the brain. In this study, we evaluated a resting-state fMRI dataset that included individuals with psychotic disorders (PD). We first estimated dynamic triple interactions using resting-state fMRI. After clustering, we estimated cohort-specific and cohort-common states for controls (CN), schizophrenia (SZ), and schizoaffective disorder (SAD). From the cohort-specific states, we observed significant triple interactions, particularly among visual, subcortical, and somatomotor networks, as well as temporal and higher cognitive networks in SZ. In SAD, key interactions involved temporal networks in the initial state and somatomotor networks in subsequent states. From the cohort-common states, we observed that high-cognitive networks were primarily involved in SZ and SAD compared to CN. Furthermore, the most significant differences between SZ and SAD also existed in high-cognitive networks. In summary, we studied PD using dynamic triple interaction, the first time such an approach has been used to study PD. Our findings highlighted the significant potential of dynamic high-order functional connectivity, paving the way for new avenues in the study of the healthy and disordered human brain.
\end{abstract}


\section{Introduction}
\label{sec:intro}
To capture temporal changes in brain activity, dynamic functional connectivity is often employed, which provides insights into how the relationships between different brain regions fluctuate during resting states or various cognitive tasks~\cite{Allen11cc}. This approach, compared to static functional connectivity, offers a more nuanced understanding of brain function. 

The traditional dynamic functional connectivity approach usually employs a sliding window method to segment blood-oxygenation-level-dependent (BOLD) signals into windows, after which it estimates the correlation matrix for each window. However, such an approach focused on pairwise relationships. Here we recognize that information interactions among brain regions are not isolated and extend beyond simple pairwise interactions~\cite{QiangEntr22,QiangNC23}. The estimation of pairwise correlations often overlooks these broader interactions, which may play a crucial role in information exchange within the brain and could serve as important biomarkers for certain psychiatric disorders. In this study, we aim to overcome this limitation and explore dynamic triple interactions in the brains of individuals with psychosis. 

\begin{figure*}[!ht]
    \centering
    \includegraphics[width=0.96\textwidth, height=9.1cm]{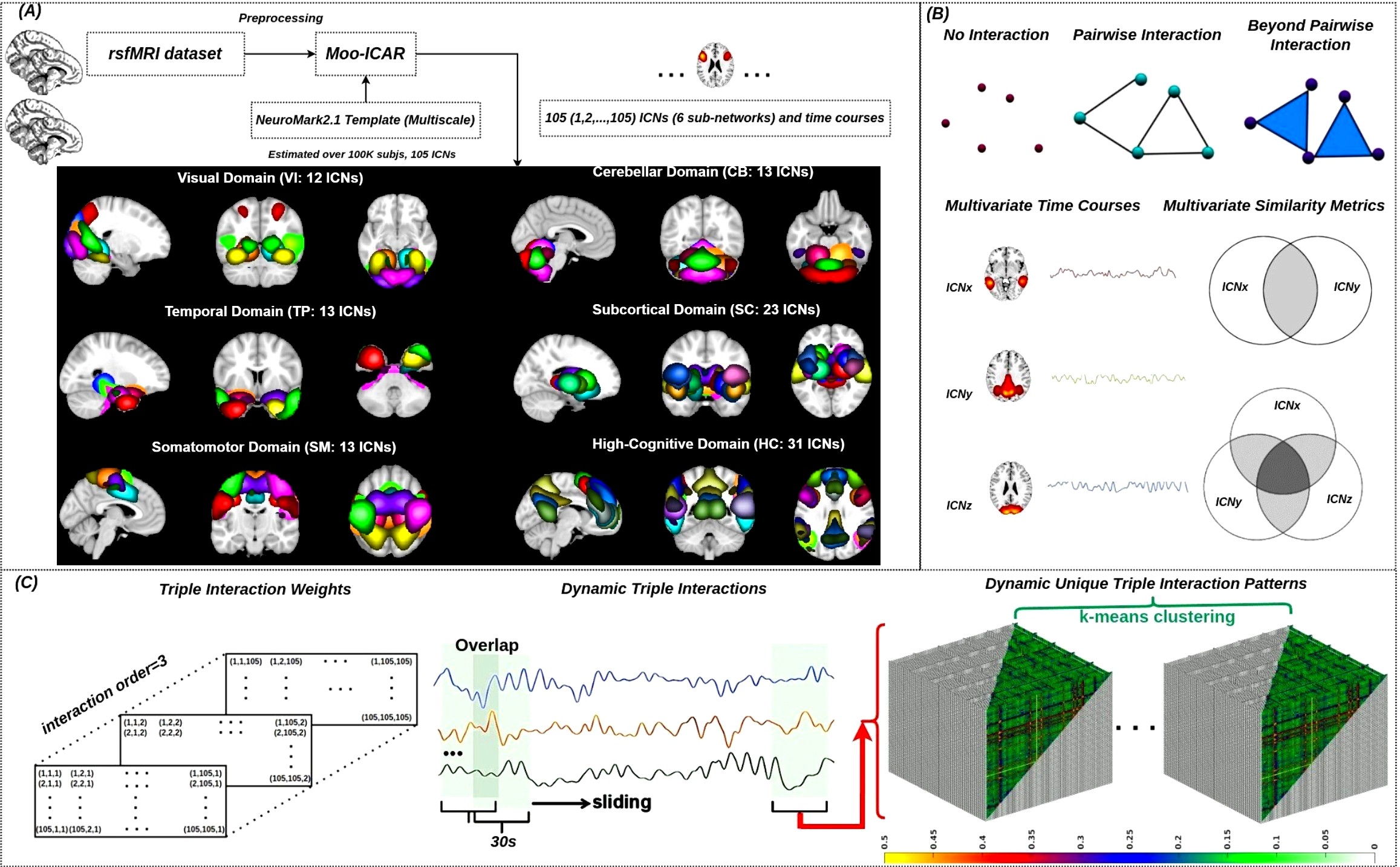}
    \caption{\textbf{Dynamic Triple Interactions Estimation Pipeline.} The preprocessed resting-state fMRI (rsfMRI) data from individuals with psychosis were input into Multi-Objective Optimization ICA with Reference (MOO-ICAR) using the spatially constrained \textit{NeuroMark2.1 Template}. This process yielded subject-specific estimates of 105 intrinsic connectivity networks (ICNs) and their corresponding time courses in psychosis rsfMRI. The 105 ICNs were then grouped into six brain sub-networks: visual (VI), cerebellar (CB), temporal (TP), subcortical (SC), somatomotor (SM), and higher cognitive (HC), as depicted in \textbf{\textit{A}}. Then the dynamic triple interactions (interaction order=3) were estimated among ICNs, as shown in \textbf{\textit{B,C}}. To estimate dynamic triple interactions, we employed a sliding window approach to segment the time series for each ICN. The triple interactions were then estimated in each window for each subject, and one of the estimated dynamic unique triple interaction patterns from a segmented window was visualized, as illustrated in \textbf{\textit{C}}. Following the clustering of all dynamic triple unique interaction patterns, we aimed to identify different states and uncover distinct patterns associated with psychosis disorder.}
    \label{fig:1}
\end{figure*}

\section{Methodology}
\label{sec:meths}
\subsection{Data Information and Preprocessing}
We analyzed resting-state fMRI data from the Bipolar and Schizophrenia Network for Intermediate Phenotypes (BSNIP). Additional information about the data and acquisition parameters can be found in~\cite{Tamminga14}. To consider computational cost and memory (the reasons can be found in Section 2.3), we used data from 166 subjects, including 59 controls (CN), 59 individuals with schizophrenia (SZ), and 48 individuals with schizoaffective disorder (SAD). Each subject's data was corrected for rigid body motion and slice timing, warped to the MNI template using nonlinear registration, and spatially smoothed with a Gaussian kernel at a full width at half maximum of 6 mm.

\subsection{Estimating Subject-Specific ICNs and Their Time Courses in the Psychosis Dataset}
\begin{figure}[!bp]
    \centering
    \includegraphics[width=0.48\textwidth, height=11.7cm]{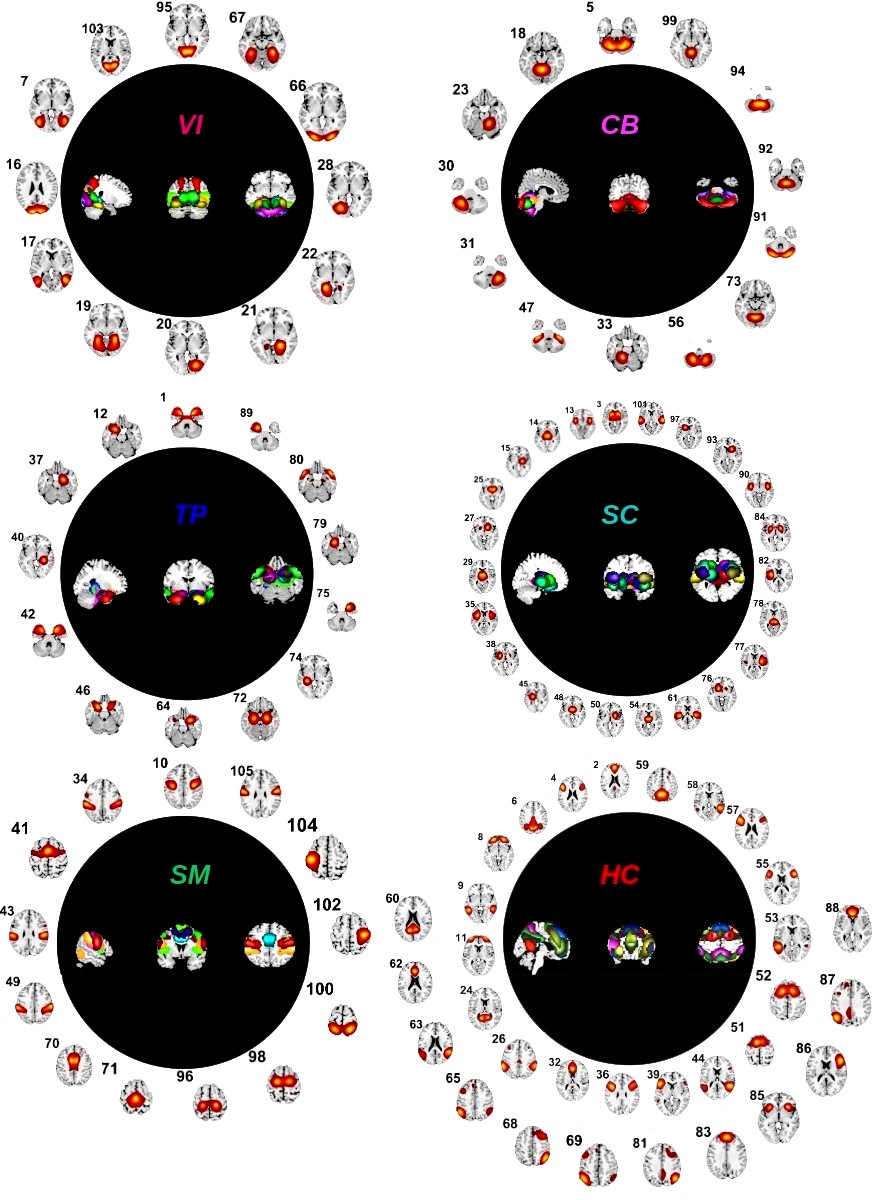}
    \caption{\textbf{Neuromark\_fMRI\_2.1 Network Template.} The \textit{Neuromark\_fMRI\_2.1 network template}~\cite{Iraji2022CanonicalAR} is available at \url{https://trendscenter.org/data/.} It includes a total of 105 intrinsic connectivity networks (ICNs) across various domains: VI (12 ICNs), CB (13 ICNs), TP (13 ICNs), SC (23 ICNs), SM (13 ICNs), and HC (31 ICNs).}
    \label{fig:2}
\end{figure}

We used multi-objective optimization ICA with reference (MOO-ICAR) to estimate subject-specific ICNs and their time courses using the \textit{Neuromark\_fMRI\_2.1\_modelorder-multi template}~\cite{Iraji2022CanonicalAR}, as depicted in Figs.\ref{fig:1}, \ref{fig:2}. We employed the Group ICA of fMRI Toolbox (GIFT) v4.0c package (\url{http://trendscenter.org/software/gift}) to implement this technique~\cite{Iraji2020ToolsOT}. Here, MOO-ICAR is a spatially constrained ICA (scICA) method that has been shown to effectively capture sample-specific information across varying data lengths and ICNs.

\subsection{Estimating Dynamic Triple Functional Connectivity}
Here, we employed Total Correlation (TC) to estimate triple interactions. The TC describes the dependence among $n$ variables ($X^1, \cdots, X^{n}$) and can be considered as a non-negative generalization of the concept of mutual information from two parties to $n$ parties. Let the definition of total correlation from Watanabe~\cite{Watanabe60} be denoted as:

\begin{equation}\label{eq.tc}
\begin{split}
   & TC\left(X^1, \cdots, X^{n}\right) = \sum_{i=1}^n h\left(X^{i}\right)-h\left(X^1, \cdots, X^{n}\right) \end{split}
\end{equation}
Where $h\left(X^{i}\right)$ is marginal entropy, and $h\left(X^1, \cdots, X^{n}\right)$ is joint entropy. The TC will be equal to mutual information if we only have two variables, and TC will be zero if all variables are totally independent.

In real situations, estimating the marginal entropy $h\left(X^i\right)$ is straightforward, but estimating the joint entropy $h\left(X^1, X^2 \cdots, X^n\right)$ is considerably challenging. To address this challenge, Gaussian information theory is commonly applied to estimate total correlation because the BOLD signals satisfy Gaussian distributions~\cite{QiangEntr22,QiangNC23}. For a univariate Gaussian random variable $X \sim \mathcal{N}(\mu, \sigma)$, the entropy (given in nats) will be,

\begin{equation}
    h^{\mathcal{N}}(X)=\frac{\ln \left(2 \pi e \sigma^2\right)}{2}
\end{equation}

And for a multivariate Gaussian distribution, the joint entropy will be,

\begin{equation}
h^{\mathcal{N}}\left(X^1, \cdots, X^n\right)=\frac{\ln \left[(2 \pi e)^n|\Sigma|\right]}{2}    
\end{equation}

The Gaussian estimator for TC can be:

\begin{equation}
    TC^{\mathcal{N}}\left(X^1, \cdots, X^n\right)=\frac{-\ln (|\Sigma|)}{2}
    \label{Eq.gtc}
\end{equation}

where \(|\Sigma|\) refers to the determinant of the covariance matrix of \(\left(X^1, \cdots, X^n\right)\). Now, we can estimate dynamic interactions beyond two-way by computing \(\vect{k}\)-way interactions (\(\vect{k} = 3\)) among \(\vect{n} = 105\) ICNs, iterating over each set of windows per subject used to obtain TC. In this study, the tapered window was created by convolving a rectangular window (width = 30 seconds) with a Gaussian function (\(\vect{\sigma} = 3\) seconds), using a sliding step size of one~\cite{Allen11cc}. In total, there are \(105^{3} = 1,157,625\) triple interactions for each window in each subject when considering \(\vect{k} = 3\) for all possible triple interactions, and there are \(\binom{105}{3} = 187,460\) unique sets of triple interactions for each window in each subject for multiscale psychotic brain networks.

\subsection{Estimating Cohort-Specific and Cohort-Common States} 
To estimate cohort-specific states for CN, SZ, and SAD, we applied k-means clustering using squared Euclidean distance. This process involved 200 iterations and was repeated five times to identify distinct brain states for each group. The optimal number of clusters (K=5) was selected using the elbow criterion. K-means clustering was performed separately for the CN, SZ, and SAD cohorts.

To estimate cohort-common states for CN, SZ, and SAD, we concatenate all subjects and repeat the same analysis used for cohort-specific states. This analysis can help us conduct group comparisons to identify connectivity differences among CN, SZ, and SAD.

\section{Results}
\label{sec:res}
\subsection{Cohort-Specific States in CN, SZ, and SAD}
\begin{figure*}[!htbp]
    \centering
    \includegraphics[width=0.95\textwidth, height=7.1cm]{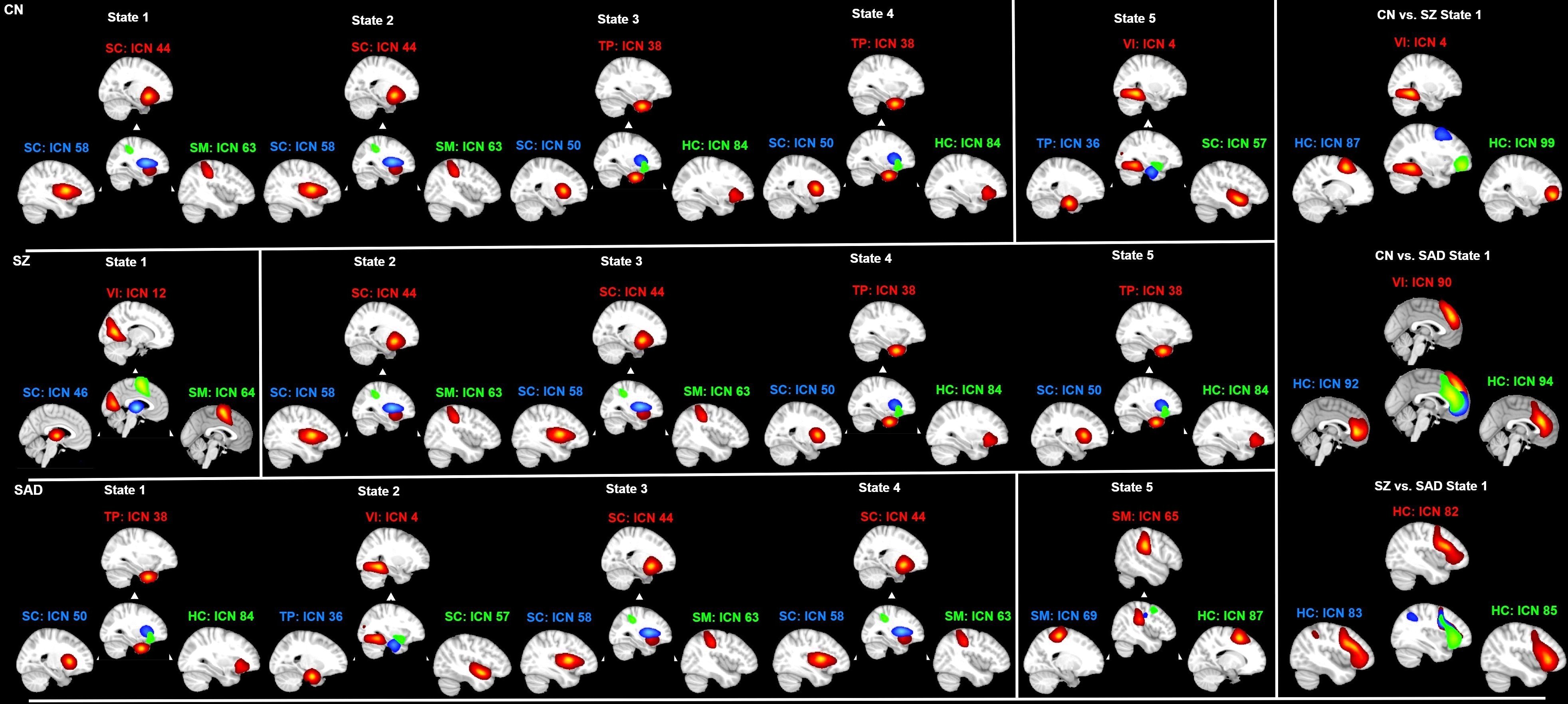}
    \caption{\textbf{The identified triple interactions in cohort-specific and cohort-common states for CN, SZ, and SAD.} The left panel shows the identified strongest triple interactions from cohort-specific states in CN, SZ, and SAD. The right panel shows the most significant triple interactions identified from cohort-common state in CN, SZ, and SAD. More explanation can be found in the results section.}
    \label{fig:3}
\end{figure*}

A total of 5 states were identified in CN, SZ, and SAD after conducting k-means clustering on the dynamic triple interaction networks, as illustrated in the left panel of Fig.\ref{fig:3}. In CN, we selected the strongest triple networks within each state. For instance, we observed that the SC-SM network interaction was prominent in State 1 and State 2. Additionally, the TP-SC-HC network interactions were present in State 3 and State 4. Finally, the VI-TP-SC interaction was notably significant in State 5. 

In SZ, we observed that the VI-SC-SM network interaction was prominent in state 1. Additionally, the SC-SM network interactions were significant in state 2 and state 3. Finally, the TP-SC-HC interaction was notably significant in state 4 and state 5. 

In SAD, we noticed that the TP-SC-HC network interaction was present in state 1, while the VI-TP-SC network interactions were observed in state 2. The SC-SM network interaction existed in state 3 and state 4. Finally, we noted that the SM-HC network interaction appeared in state 5. 

\subsection{Cohort-Common States in CN, SZ, and SAD}

For group comparisons, we identified 5 states and conducted statistical comparisons across all of them. We then performed a two-sample t-test between CN vs. SZ, CN vs. SAD, and SZ vs. SAD in State 1, as it exhibited the strongest connectivity compared to the other states. Our analysis revealed that the most significant networks involved in SZ and SAD were high-cognitive networks when compared to CN. We observed that the major differences between SZ and SAD were primarily located in high-cognitive networks. We then selected the state one with the most differences to highlight, as shown in the right panel of Fig.\ref{fig:3}.

In summary, we utilized dynamic triple interaction to improve our understanding of the psychotic brain. Furthermore, our findings demonstrated the significant potential of dynamic high-order functional connectivity, opening new pathways for research on the psychotic brain. 

\section{Future Work}
\label{sec:fw}
In the future, quantifying dynamic triple interactions will enhance our understanding of how states switch and relate. It will also reveal abnormal network interactions and help identify patterns linked to psychiatric conditions. Moreover, using a more refined and precise Neuromark network template (e.g. \textit{Neuromark\_fMRI\_2.2 Network Template}) will allow us to explore intricate network interactions in psychiatric studies with enhanced precision. In addition, a more in-depth analysis of group comparisons among CN, SZ, and SAD, as well as comparisons between dynamic triple and pairwise interactions, can be pursued in the future.

\section{Acknowledgments}
The authors declare that there are no conflicts of interest related to this research. This work was supported by NSF grant 2112455, and NIH grants R01MH123610 and R01MH119251.
\bibliographystyle{ieeetr}
\bibliography{references} 

\begin{thebibliography}{1}

\bibitem{Allen11cc}
E.~Allen, E.~Damaraju, S.~Plis, E.~Erhardt, T.~Eichele, and V.~Calhoun, ``Tracking whole-brain connectivity dynamics in the resting state,'' {\em Cerebral cortex (New York, N.Y. : 1991)}, 11 2012.

\bibitem{QiangEntr22}
Q.~Li, G.~V. Steeg, S.~Yu, and J.~Malo, ``Functional connectome of the human brain with total correlation,'' {\em Entropy}, vol.~24, no.~12, 2022.

\bibitem{QiangNC23}
Q.~Li, G.~Ver~Steeg, and J.~Malo, ``Functional connectivity via total correlation: Analytical results in visual areas,'' {\em Neurocomputing}, vol.~571, p.~127143, 12 2023.

\bibitem{Tamminga14}
C.~A. Tamminga, G.~Pearlson, M.~Keshavan, J.~Sweeney, B.~Clementz, and G.~Thaker, ``Bipolar and schizophrenia network for intermediate phenotypes: Outcomes across the psychosis continuum,'' {\em Schizophrenia Bulletin}, vol.~40, pp.~S131--S137, 02 2014.

\bibitem{Iraji2022CanonicalAR}
A.~Iraji, Z.~Fu, A.~Faghiri, M.~Duda, J.~Chen, S.~Rachakonda, T.~DeRamus, P.~Kochunov, B.~Adhikari, A.~Belger, J.~Ford, D.~Mathalon, G.~Pearlson, S.~Potkin, A.~Preda, J.~Turner, T.~Erp, J.~Bustillo, K.~Yang, and V.~Calhoun, ``Identifying canonical and replicable multi‐scale intrinsic connectivity networks in 100k+ resting‐state fmri datasets,'' {\em Human Brain Mapping}, vol.~44, 10 2023.

\bibitem{Iraji2020ToolsOT}
A.~Iraji, A.~Faghiri, N.~Lewis, Z.~Fu, S.~Rachakonda, and V.~D. Calhoun, ``Tools of the trade: estimating time-varying connectivity patterns from fmri data,'' {\em Social Cognitive and Affective Neuroscience}, vol.~16, pp.~849 -- 874, 2020.

\bibitem{Watanabe60}
S.~Watanabe, ``Information theoretical analysis of multivariate correlation,'' {\em IBM Journal of research and development}, vol.~4, no.~1, pp.~66--82, 1960.

\end{thebibliography}

\end{document}